\documentclass[aps,prd,preprintnumbers,showpacs,showkeys,nofootinbib,
superscriptaddress,fleqn,floatfix,tightenlines,10pt]{revtex4-1}
\usepackage{amsmath,amsfonts,amssymb,amscd,amsxtra,amsthm}
\usepackage{graphicx}  
\usepackage{epstopdf}
\usepackage{dcolumn}  
\usepackage{bm}          
\usepackage{slashed}
\usepackage{cancel}
\usepackage{float}
\usepackage{mathtools}
\usepackage{amsbsy}
\usepackage{amstext}

\usepackage[utf8]{inputenc} 
\usepackage{booktabs}
\usepackage[normalem]{ulem} 
\usepackage[dvipsnames]{xcolor} 
\usepackage{tabularx}
\usepackage{enumitem}  
\usepackage{array} 
\usepackage{slashed}
\usepackage{tikz}
\usepackage{cleveref} 
\usepackage{multirow}
\renewcommand\sout{\bgroup \color{red} \ULdepth=-.5ex \ULset}

\makeatletter

\begin{document}  
\preprint{INHA-NTG-14/2018}
\title{Electromagnetic properties of singly heavy baryons}
\author{June-Young Kim}
\email[E-mail: ]{junyoung.kim@inha.edu}
\affiliation{Department of Physics, Inha University, Incheon 22212,
Republic of Korea}

\author{Hyun-Chul Kim}
\email[E-mail: ]{hchkim@inha.ac.kr}
\affiliation{Department of Physics, Inha University, Incheon 22212,
Republic of Korea}
\affiliation{School of Physics, Korea Institute for Advanced Study 
  (KIAS), Seoul 02455, Republic of Korea}
\affiliation{  Advanced Science Research Center, Japan Atomic 
Energy Agency, Shirakata, Tokai, Ibaraki, 319-1195, Japan}

\date{\today}
\begin{abstract}
In this presentation, we summarize selectively recent results of the 
electromagnetic form factors of singly heavy baryons with spin 1/2 and
related quantities, which were derived within a framework of the 
self-consistent chiral quark-soliton model. The results are compared
with those from the lattice QCD and their physical implications are
discussed.  
\end{abstract}
\maketitle

\section{Introduction}
The chiral quark-soliton model ($\chi$QSM) was developed as a pion
mean-field approach to describe the structures of the
nucleon~\cite{Diakonov:1987ty,Diakonov:1997sj}. The model was
extended to describe also 
hyperons~\cite{Blotz:1992pw,Praszalowicz:1998jm,Christov:1995vm}. Recently,
the model was 
successfully applied to the description of singly heavy baryons
\cite{Yang:2016qdz,Kim:2017jpx,Kim:2017khv,Kim:2018xlc,Yang:2018uoj}
in the limit of the infinitely heavy quark mass ($m_Q\to\infty$).    
In this presentation, we summarize selectively a recent work on the 
electromagnetic form factors of singly heavy baryons with spin
1/2~\cite{Kim:2018nqf} and compare the results with those from a
lattice QCD~\cite{Can:2013tna}. 
\section{Results and discussion}
\begin{figure}[htp]
\centering
\includegraphics[scale=0.285]{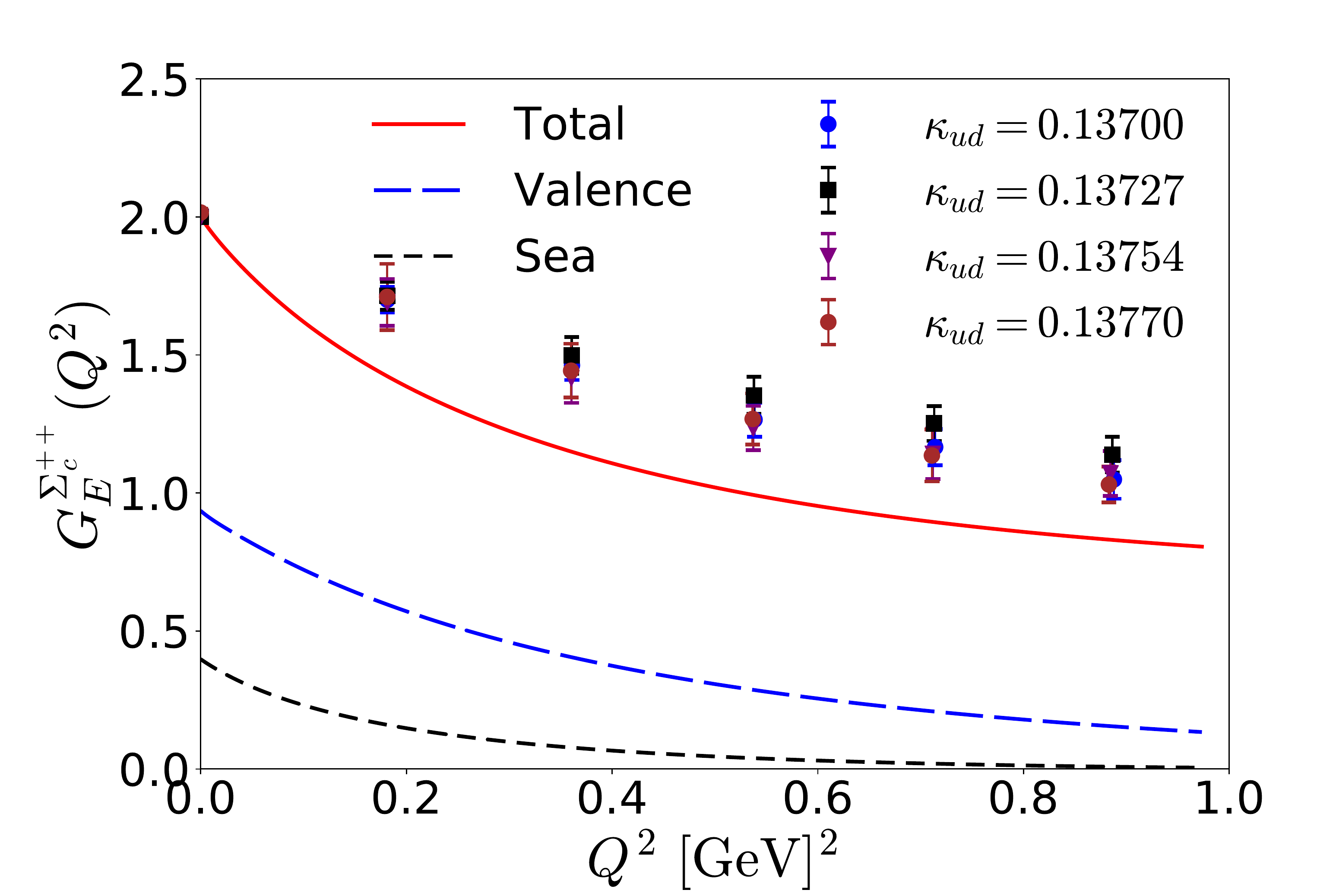}
\includegraphics[scale=0.285]{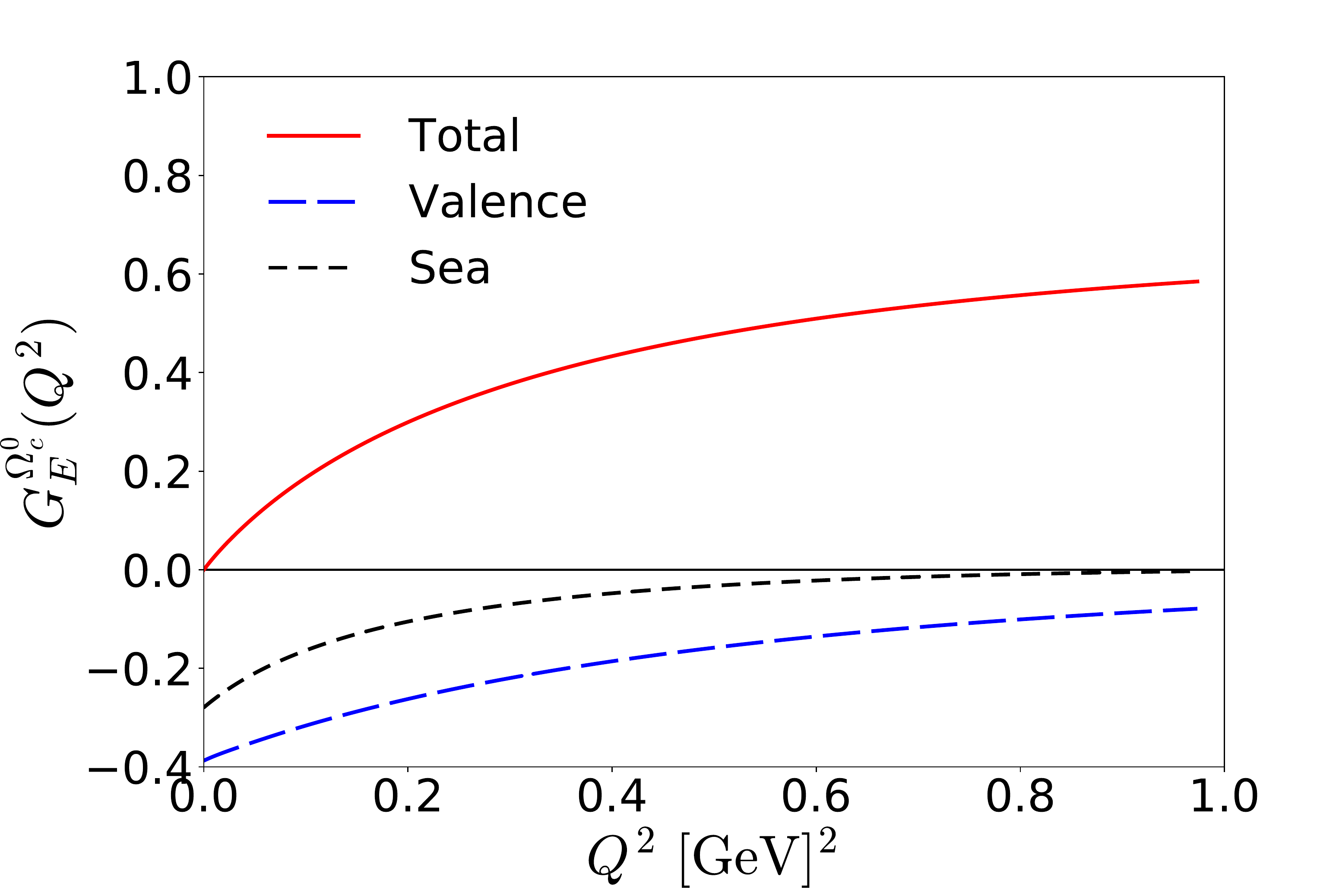}
\caption{Decomposition of the valence and sea contributions of the
  electric form factors for the charmed baryons $\Sigma_c^{++}$ and
  $\Omega_c^0$, in the left panel,  
  and the right panel, respectively. The results are compared with
  those from the lattice QCD.} 
\label{fig:1}
\end{figure}
\begin{figure}[htp]
\centering
\includegraphics[scale=0.285]{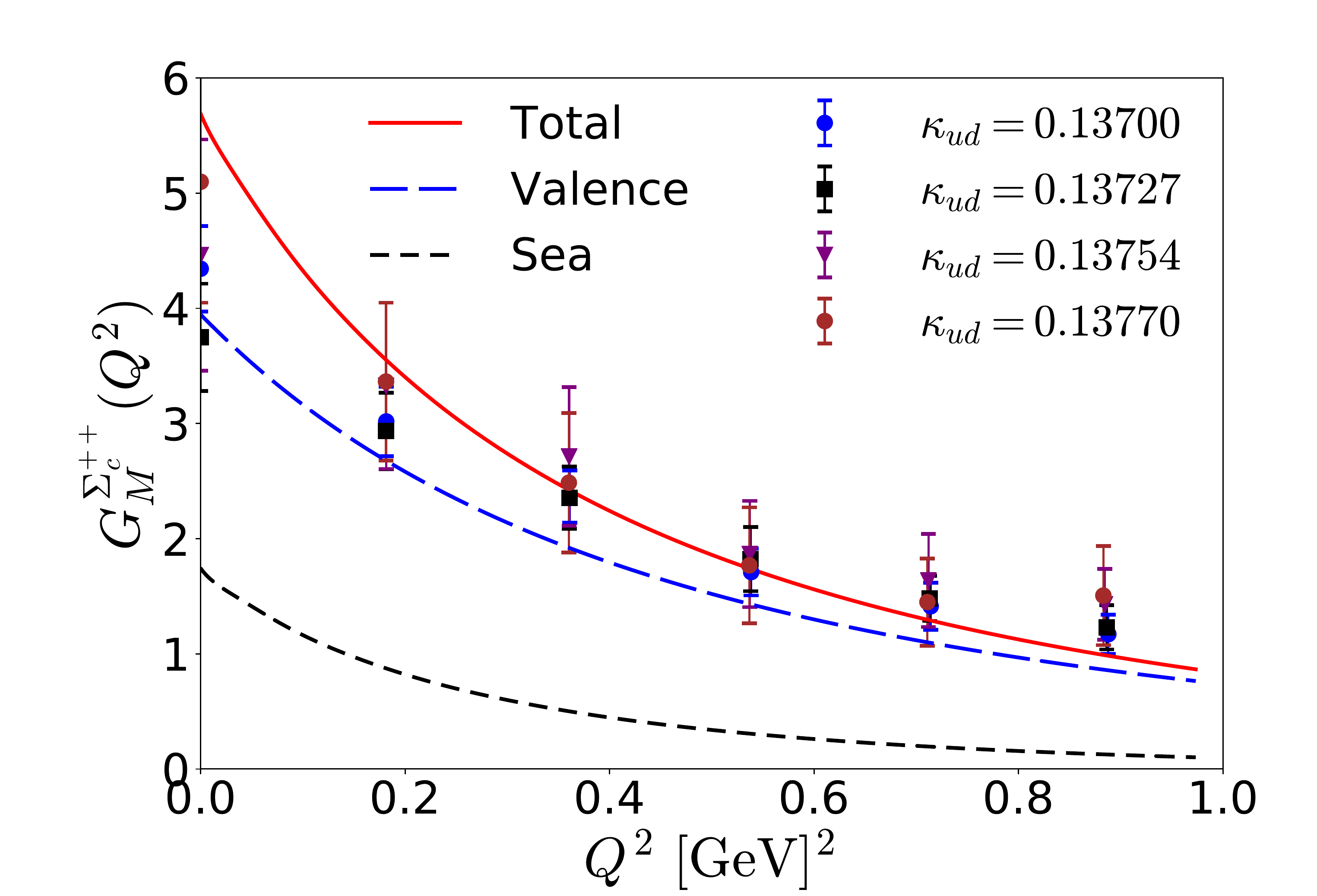}
\includegraphics[scale=0.285]{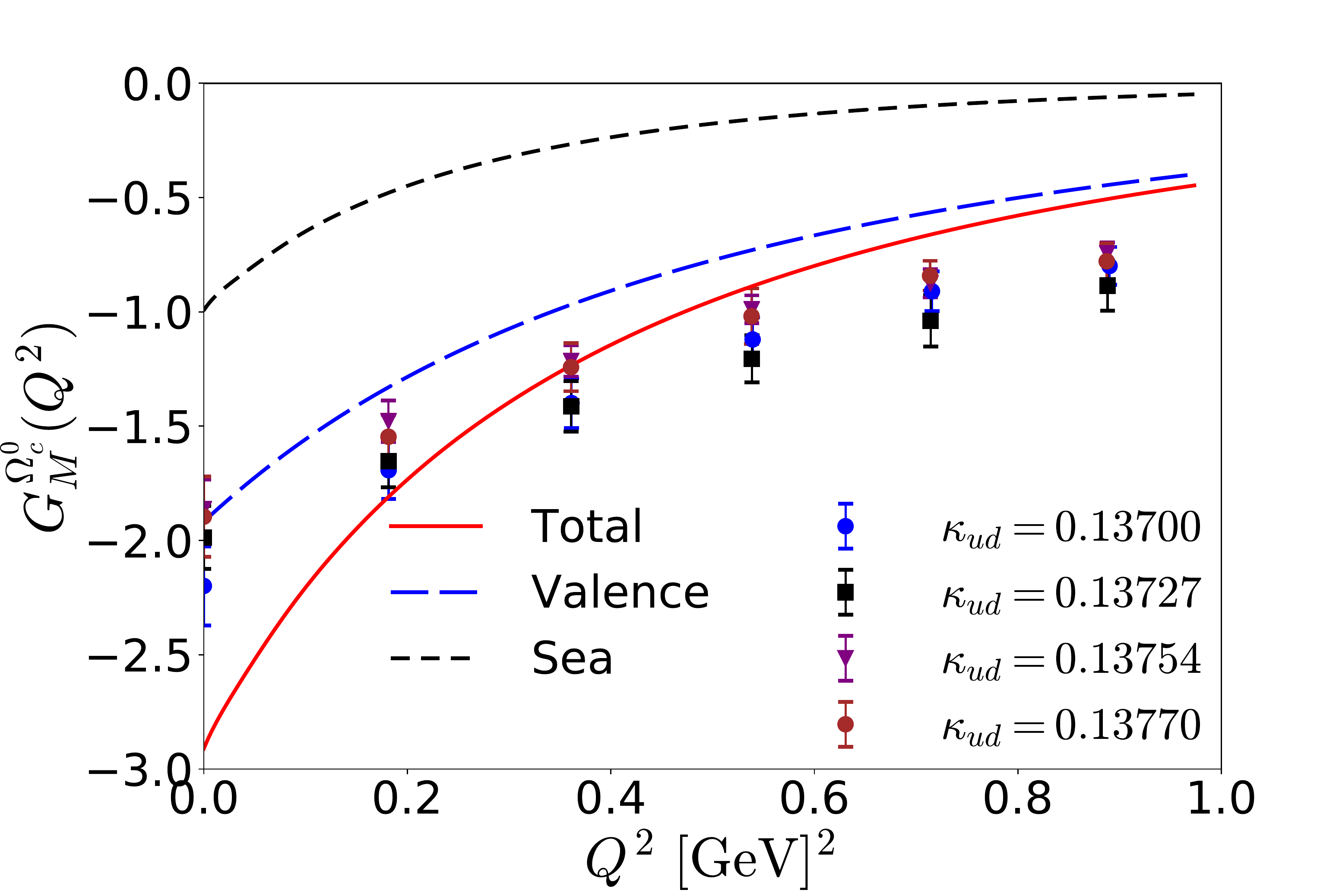}
\caption{Decomposition of the valence and sea contributions of the
  magnetic form factors for the charmed baryons $\Sigma_c^{++}$ and
  $\Omega_c^0$, in the left panel,  
  and the right panel, respectively. The results are compared with
  those from the lattice QCD. }
\label{fig:2}
\end{figure}

In Fig.~\ref{fig:1}, we show the results of the
electric form factors of the charmed heavy baryons $\Sigma_c^{++}$ and
$\Omega_c^0$, comparing them with those from the lattice
QCD~\cite{Can:2013tna}.  
The present results seem to fall off faster than those from the
lattice QCD. This can be easily understood: When the value of the pion
mass employed by the lattice QCD calculation is larger than the
experimental one, the nucleon EM form factors decrease less slowly
than the experimental data. This tendency can be also found in the
magnetic form factors of the heavy baryons as drawn in Fig.~\ref{fig:2}.

\begin{table}
\caption{Dipole magnetic moments of the singly heavy baryons. The
  second and third columns list the present results without and with
  the effects of SU(3) symmetry breaking, respectively.}

\centering
\begin{tabular}{c | c c c c}
\hline
\hline 
$B_{c}$ & $\mu_{B_{c}}^{(m_{s}=0)}$ & $\mu_{B_{c}}^{(m_{s}=180)}$ &
\cite{Can:2013tna,Bahtiyar:2016dom} & ~\cite{Yang:2018uoj}\\ 
\hline
$\Sigma^{++}_{c}$&2.147 & 2.176 & 2.220(505)& 2.15(10)\\
$\Sigma^{+}_{c}$ &0.537& 0.445 & - & 0.46(3)\\
$\Sigma^{0}_{c}$ &-1.073& -1.286& -1.073(269) & -1.24(5)\\
$\Xi^{+}_{c}$&0.537 & 0.603 & 0.315(141) & 0.60(2) \\
$\Xi^{0}_{c}$&-1.073 &-1.147 &-0.599(71) & -1.05(4)\\
$\Omega^{0}_{c}$ &-1.073 &  -1.014 & -0.639(88) & -0.85(5)\\
\hline
\hline
\end{tabular}
\label{tab:1}
\end{table}
In Fig.~\ref{fig:1}, we depict the light-quark contributions of 
the valence and sea parts to the electric form factors
separately. In the present pion mean-field approach and in the limit
of $m_Q\to\infty$, the heavy quark can be considered as a point-like 
particle, so that it plays only a part in making the concerned baryon
charge complete. Note that the heavy-quark contribution to the
magnetic form factors also vanishes in the limit of
$m_Q\to\infty$. So, the light quarks govern all dynamics in a heavy
baryon. 

In Table~\ref{tab:1}, we list the results of the magnetic moments of
the charmed baryons, comparing them with those from the lattice QCD
and from a ``model-independent'' approach of the $\chi$QSM. We want to
mention that the nuclear magneton in the model is evaluated not by the
experimental value of the nucleon mass but by the model value, which
is larger than the experimental one. Keeping this in mind, we find
that except for $\mu_{\Sigma^{++}}$ the present results turn out
larger than those from the lattice QCD.

\vspace{0.5cm}
The work is supported by Basic Science Research Program through the
National Research Foundation (NRF) of Korea funded by the 
Korean government (Ministry of Education, Science and
Technology(MEST)): Grant No. NRF-2018R1A2B2001752. 
%


\begin{thebibliography}{99}
\bibitem{Diakonov:1987ty} 
  D.~Diakonov, V.~Y.~Petrov and P.~V.~Pobylitsa,
  Nucl.\ Phys.\ B {\bf 306}, 809 (1988).

\bibitem{Diakonov:1997sj} 
  D.~Diakonov,
  hep-ph/9802298.

\bibitem{Blotz:1992pw} 
  A.~Blotz, D.~Diakonov, K.~Goeke, N.~W.~Park, V.~Petrov and P.~V.~Pobylitsa,
  Nucl.\ Phys.\ A {\bf 555}, 765 (1993).

\bibitem{Praszalowicz:1998jm} 
  M.~Praszalowicz, T.~Watabe and K.~Goeke,
  Nucl.\ Phys.\ A {\bf 647}, 49 (1999).

\bibitem{Christov:1995vm} 
  C.~V.~Christov, A.~Blotz, H.-Ch.~Kim, P.~Pobylitsa, T.~Watabe,
  T.~Meissner, E.~Ruiz Arriola and K.~Goeke, 
  Prog.\ Part.\ Nucl.\ Phys.\  {\bf 37}, 91 (1996).

\bibitem{Yang:2016qdz} 
  Gh.-S.~Yang, H.-Ch.~Kim, M.~V.~Polyakov and M.~Prasza{\l}owicz,
  Phys.\ Rev.\ D {\bf 94}, 071502 (2016).

\bibitem{Kim:2017jpx} 
  H.-Ch.~Kim, M.~V.~Polyakov and M.~Prasza{\l}owicz,
  Phys.\ Rev.\ D {\bf 96}, 014009 (2017)
  Addendum: [Phys.\ Rev.\ D {\bf 96}, 039902 (2017)].

\bibitem{Kim:2017khv} 
  H.-Ch.~Kim, M.~V.~Polyakov, M.~Praszalowicz and G.~S.~Yang,
  Phys.\ Rev.\ D {\bf 96}, 094021 (2017)
  Erratum: [Phys.\ Rev.\ D {\bf 97}, 039901 (2018)].

\bibitem{Kim:2018xlc} 
  J.~Y.~Kim, H.-Ch.~Kim and G.~S.~Yang,
  Phys.\ Rev.\ D {\bf 98}, 054004 (2018).

\bibitem{Yang:2018uoj} 
  G.~S.~Yang and H.-Ch.~Kim,
  Phys.\ Lett.\ B {\bf 781}, 601 (2018).


\bibitem{Kim:2018nqf} 
  J.~Y.~Kim and H.-Ch.~Kim,
  Phys.\ Rev.\ D {\bf 97}, 114009 (2018).

\bibitem{Can:2013tna} 
  K.~U.~Can, G.~Erkol, B.~Isildak, M.~Oka and T.~T.~Takahashi,
  JHEP {\bf 1405}, 125 (2014).

\bibitem{Bahtiyar:2016dom}
  H.~Bahtiyar, K.~U.~Can, G.~Erkol, M.~Oka and T.~T.~Takahashi,
  Phys.\ Lett.\ B {\bf 772} (2017) 121
  doi:10.1016/j.physletb.2017.06.022
  [arXiv:1612.05722 [hep-lat]].
\end{thebibliography}
\end{document}